\documentclass[12pt]{article}

\advance \textheight by \topskip
\makeatletter
\@addtoreset{equation}{section}
 
\makeatother

\def\R{\mathbb R}

\def\N{\mathbb N}
\def\Z{\mathbb Z}

\usepackage{amsmath,amssymb}

\begin{document}
\title{
{\bf
Interaction via Reduction and
Nonlinear Superconformal Symmetry
}}

\author{{\sf Andres Anabalon${}^{a}$}
{\sf\ and Mikhail S. Plyushchay${}^{a,b}$}\thanks{
E-mail: mplyushc@lauca.usach.cl}
\\
{\small {\it ${}^a$Departamento de F\'{\i}sica,
Universidad de Santiago de Chile,
Casilla 307, Santiago 2, Chile}}\\
{\small {\it ${}^b$Institute for High Energy Physics,
Protvino, Russia}}}
\date{}






\maketitle


\begin{abstract}
We show that the reduction of a planar
free spin-$\frac{1}{2}$ particle
system by the constraint fixing its total angular momentum
produces the one-dimensional
Akulov-Pashnev-Fubini-Rabinovici
superconformal mechanics model
with the nontrivially coupled boson and fermion
degrees of freedom.
The modification of the constraint
by including the particle's spin with
the relative weight $n\in \N$, $n>1$,
and subsequent application of the
Dirac reduction procedure
(`first quantize and then reduce')
give rise to the anomaly free quantum system
with the order $n$ nonlinear superconformal symmetry
constructed recently in hep-th/0304257.
We  establish the origin of the
quantum corrections to the integrals of motion
generating the nonlinear superconformal algebra,
and fix completely its
form.
\end{abstract}

\newpage
\section{Introduction}

The superconformal mechanics
was introduced twenty years ago
by Akulov and Pashnev \cite{AP},
and by Fubini and Rabinovici \cite{FR}
as a supersymmetric analog of the conformal
mechanics of De Alfaro, Fubini and Furlan \cite{AFF}.
It was examined in different aspects
in  \cite{IKL,FM},
and nowadays the interest to
the conformal and superconformal
mechanics is related mainly to the AdS/CFT
correspondence conjecture
and to the integrable models
\cite{BH}-\cite{BGK}.

Recently,  the
superconformal mechanics model
\cite{AP,FR}
was generalized in \cite{LP2}
following the ideas of
nonlinear supersymmetry \cite{AIS}-\cite{AS}.
In comparison with the original model
\cite{AP,FR}, the model \cite{LP2} is
characterized at the classical level
by the $n$-fold fermion-boson coupling
constant,
that gives rise to a radical change of the symmetry:
instead of the $osp(2|2)$ Lie superalgebraic structure,
the modified system possesses the order
$n$ nonlinear superconformal symmetry.
The latter is generated by the boson
integrals which form,
as in the linear $osp(2|2)$ case,
the Lie subalgebra $so(1,2)\oplus u(1)$,
while the set of $2(n+1)$ fermion integrals
of motion anticommutes for the order $n$
polynomials of the even integrals.
The two essential moments of the construction
left, however, unclarified.
First, though the quantum analogs of the
integrals of motion were found from the
requirement of preservation of the symmetry at the quantum
level,
the origin  of the quantum corrections to the integrals
remained to be completely unclear.
Second, the quantum analog
of the classical nonlinear superconformal symmetry algebra
was determined entirely
only for the simplest case of $n=2$,
whereas for the general case of  $n\in \N$
only a part of the anticommutators
of the odd integrals was fixed.

In the present paper, we shall clarify the
origin of the quantum corrections to the integrals
of motion of the model \cite{LP2}, and shall fix
completely the
form of the quantum nonlinear superconformal
algebra
of an arbitrary order $n$.
This will be done by the reduction of a
planar free  spin-$\frac{1}{2}$ particle system
to the surface of the constraint fixing linearly
the orbital angular momentum in terms of
the spin.

The paper is organized as follows.
In Section 2 we first show how the
one-dimensional model \cite{AP,FR} with the nontrivial
boson-fermion coupling
can be obtained via reduction
from the planar system of a free nonrelativistic
particle in which
spin and
translation degrees of freedom
are completely decoupled.
In Section 3 we consider a
modification of the reduction procedure
which leads to the generalized model \cite{LP2},
and fix the form of the order $n$ nonlinear
superconformal algebra.
In section  4 we shortly summarize the results
and discuss possible applications
and generalizations
of the proposed method of introducing the
boson-fermion interaction via a
reduction procedure.

\section{Superconformal symmetry: linear case}

Let us consider a free nonrelativistic
spin-$\frac{1}{2}$ particle
on the plane ($i=1,2$),
\begin{equation}
A=\int {\cal L}_0dt,\qquad
{\cal L}_0=\frac{1}{2}\dot{x}_i^2-\frac{i}{2}\dot{\xi}_i
\xi_i.
\label{act}
\end{equation}
Its Hamiltonian coincides with that of a free
2D nonrelativistic spinless particle,
$
H=\frac{1}{2}p_i^2,
$
and  so,
\begin{equation}
I=(X_i\equiv x_i-p_it,\,\, p_i,\,\,  \xi_i)
\label{int1}
\end{equation}
is the set of the integrals of motion,
\begin{equation}
\frac{d}{dt}I=\frac{\partial }{\partial t}I+
\{I,H\}=0,
\label{integr}
\end{equation}
linear
in the phase space variables of the system.
The integrals (\ref{int1}) form a
superextended 2D Heisenberg algebra,
$\{X_i,p_j\}=\delta_{ij}$, $\{\xi_i,\xi_j\}=-i\delta_{ij}$.
The even,
\begin{equation}
H=\frac{1}{2}p_i^2,\qquad
K=\frac{1}{2}X_i^2,\qquad
D=\frac{1}{2}X_ip_i,
\label{hkd}
\end{equation}
\begin{equation}
L=\epsilon_{ij}X_ip_j,\qquad
\Sigma=-\frac{i}{2}\epsilon_{ij}\xi_i\xi_j,
\label{lorb}
\end{equation}
and odd,
\begin{equation}
Q_{1}=p_i\xi_i,\qquad
Q_{2}=\varepsilon _{ij}p_{i}\xi _{j},\qquad
S_{1}=X_i\xi_i,
\qquad S_{2}=\varepsilon _{ij}X_{i}\xi _{j},
\label{odd1}
\end{equation}
quadratic combinations of (\ref{int1})
generate the following $Z_2$-graded
Lie algebra (only the nontrivial Poisson bracket relations
are displayed):
\begin{equation}
\{D,H\}=H,\quad
\{D,K\}=-K,\quad
\{K,H\}=2D,
\label{dhk}
\end{equation}
\begin{equation}
\{D,Q_a\}=\frac{1}{2}Q_a,\qquad
\{D,S_a\}=-\frac{1}{2}S_a,
\label{dqs}
\end{equation}
\begin{equation}
\{H,S_a\}=-Q_a,\qquad
\{K,Q_a\}=S_a,
\label{hkqs}
\end{equation}
\begin{equation}
\{L,Q_a\}=-\epsilon_{ab}Q_b,\quad
\{L,S_a\}=-\epsilon_{ab}S_b,\quad
\{\Sigma,Q_a\}=\epsilon_{ab}Q_b,\quad
\{\Sigma,S_a\}=\epsilon_{ab}S_b,
\label{lsqs}
\end{equation}
\begin{equation}
\{Q_a,Q_b\}=-i\delta_{ab}2H,\qquad
\{S_a,S_b\}=-i\delta_{ab}2K,
\label{qqss}
\end{equation}
\begin{equation}
\{Q_a,S_b\}=-i\delta_{ab}2D -i\epsilon_{ab}(L+2\Sigma).
\label{qsdj}
\end{equation}
The total angular momentum, $J=L+\Sigma$,
commutes with all these quadratic scalar integrals, and
(\ref{dhk})--(\ref{qsdj}) is identified as the
$osp(2|2)\oplus u(1)$
superalgebra with the $u(1)$ corresponding to the centre
$J$.
The non-Abelian part of the
bosonic subalgebra $so(1,2)\oplus u(1)$
of the conformal superalgebra $osp(2|2)\cong su(1,1|1)$
is generated here by the integrals (\ref{hkd}),
and its Abelian $u(1)$ subalgebra
is associated with the linear combination $\Sigma+ J$.

The boson  and fermion degrees of freedom in the system
(\ref{act})
are completely decoupled.
The  interaction between them can be introduced without
violating the superconformal symmetry
$osp(2|2)$
in the following
manner.
Since $J$ is the centre, it can be fixed without
changing the structure of the superalgebra
(\ref{dhk})--(\ref{qsdj}).
Let us make this by introducing the classical constraint
\begin{equation}
{\cal J}_1\equiv L+\Sigma-\alpha\approx 0,
\label{j1}
\end{equation}
where $\alpha$ is a real parameter.
The physical variables are those which
commute (in the sense of the  Poisson brackets)
with the constraint, and they
can immediately be identified with the scalars.
Having in mind the quantization and
further generalization,
it is more convenient first to pass over to the
polar coordinates and then to identify
the observables.

Let us introduce the
two orthonormal vectors
\begin{equation}
n^{(1)}_i=(\cos\varphi,\sin\varphi),\qquad
n^{(2)}_i=-\epsilon_{ij}n^{(1)}_j=(-\sin\varphi,
\cos\varphi),
\label{n12}
\end{equation}
in terms of which
\begin{equation}
x_i=rn^{(1)}_i,\qquad
p_i=p_rn^{(1)}_i+r^{-1}Ln^{(2)}_i,\qquad
\xi_i=\xi^{(a)}n^{(a)}_i.
\label{xpx}
\end{equation}
The transformation
(\ref{n12}), (\ref{xpx})
to the
new variables is canonical,
\[
\{r,p_r\}=1, \qquad
\{\varphi,L\}=1,\qquad
\{\xi^{(a)},\xi^{(b)}\}=-i\delta^{ab},
\]
but it
is well defined only for
$r=\sqrt{x_i^2}>0$.
This will be essential
for the quantum theory.
The scalar variables
\begin{equation}
q\equiv r,\qquad
p\equiv p_r,\qquad
\psi_a\equiv\xi^{(a)}
\label{obs}
\end{equation}
commute with the constraint (\ref{j1}),
and any function of them is an observable.

On the surface of the
constraint (\ref{j1}), the orbital angular momentum
is given in terms of $\Sigma=-i\psi_1\psi_2$,
$
L=\alpha-\Sigma.
$
Then the reduction of the Hamiltonian
$
H=\frac{1}{2}(p_r^2+r^{-2}L^2)
$
to the surface (\ref{j1}) produces the
nontrivial boson-fermion
interaction in the resulting one-dimensional system:
\begin{equation}
H=\frac{1}{2}\left(p^2+\frac{1}{q^2}\alpha(\alpha +i\psi_1
\psi_2)\right).
\label{hconf}
\end{equation}
The Hamiltonian (\ref{hconf}) coincides with
that of the classical superconformal model
\cite{AP,FR},
and
the quantities  $H$, $K$, $D$, $\Sigma+\alpha$,
$Q_a$ and $S_a$ being rewritten
in terms of observables (\ref{obs}),
take the form  of the generators of
the $osp(2|2)$ superconformal symmetry for
the system (\ref{hconf})
(see below).

The direct quantization of the classical
one-dimensional system (\ref{hconf})
on the half-line $q>0$
reproduces the quantum superconformal model
\cite{AP,FR}. However, having in mind that
the two procedures ---
`first reduce and then quantize' and
`first quantize and then reduce' ---
generally give different results \cite{PR},
let us consider shortly the latter procedure.
It is this method that will reproduce
all the quantum corrections
in the quantum analogs of the classical integrals
necessary for preserving the nonlinear
superconformal symmetry.
In ref. \cite{LP2} the corresponding corrections
were introduced by hands, and so, their origin
was unclear.

In terms of the polar coordinates
the scalar product is
\begin{equation}
\left( \Psi _{1},\Psi _{2}\right) =\int_{0}^{\infty }
rdr\int_{0}^{2\pi
}\Psi _{1}^{\ast }(r,\varphi)
\Psi _{2}(r,\varphi)d\varphi.
\label{scprod}
\end{equation}
Here $\Psi(r,\varphi)$ is a two-component
(spinor)
wave function, on which the quantum
analogs of the Grassmann variables $\psi_a$
act as the Pauli matrices:
\[
\hat\psi_a=\sqrt{\frac{\hbar}{2}}\sigma_a,
\qquad a=1,2.
\]
In what follows we put $\hbar=1$.
With respect to the scalar product
(\ref{scprod})
the operators
\begin{equation}
\hat{p}_{r}=-i\dfrac{\partial }{\partial r}-\dfrac{i}{2r},
\qquad
\hat{L}=\epsilon_{ij}x_ip_j=-i
\dfrac{\partial }{\partial \varphi }
\end{equation}
are Hermitian.
The quantum analog of the constraint (\ref{j1}) specifies
the physical subspace of the system:
\begin{equation}
\left(\hat{L}+\frac{1}{2}\sigma_3-\alpha\right)\Psi_{phys}=
0.
\label{qj1}
\end{equation}
In (\ref{qj1}) the second term corresponds to
the particle's spin
$\hat\Sigma=-\frac{i}{2}[\hat\psi_1,\hat\psi_2]$.
Taking into account the $2\pi$-periodicity of the wave
functions,
$\Psi(r,\varphi)=\sum_{l=-\infty}^{+\infty}\Phi_l(r)
e^{il\varphi}$,
we find that eq. (\ref{qj1}) has nontrivial
solution of the form
\[
\Psi_{phys}(r,\varphi)=\left(
\begin{array}{c}
\Phi _{+}(r) \\
\Phi _{-}(r)e^{i\varphi}
\end{array}
\right)e^{ik\varphi}
\]
only when the parameter $\alpha$
takes a half-integer value,
\begin{equation}
\alpha=k+\frac{1}{2},
\label{aq}
\end{equation}
where $k$ is a fixed integer number, $k\in\Z$.
The redefinition of the
radial wave functions according to
$\Phi(r)\rightarrow
\phi(r)=\sqrt{2\pi r}\Phi(r)$,
and integration in the angular variable
reduce (\ref{scprod})
to the scalar product on the half-line,
\begin{equation}
(\phi,\phi')=\int_0^{\infty}\phi^*(q)\phi'(q)dq.
\label{newsc}
\end{equation}
The spinor wave functions $\phi(q)$  are subject now to the
boundary
condition $\phi(q)|_{q\rightarrow 0} =0$, and
the action of the
operator $\hat p_r$ is reduced on them to the operator
$
\hat{p}=-i\partial/\partial q.
$
Having in mind the quantum relations
\begin{equation}
\hat p_i^2=\hat p_r^2+\frac{1}{r^2}\left(\hat
L^2-\frac{1}{4}\right)
\label{hrad}
\end{equation}
and (\ref{qj1}), the reduced quantum
Hamiltonian takes the form of the
Hamiltonian of the one-dimensional quantum
superconformal
mechanics model \cite{AP,FR},
\begin{equation}
\hat H=\frac{1}{2}\left(
-\frac{\partial ^2}{\partial q^2}+\frac{1}{q^2}
\alpha(\alpha-\sigma_3)\right).
\label{hred1}
\end{equation}
However, in accordance with relation (\ref{aq}),
here the quantized parameter $\alpha$
takes only half-integer values.
In order it could take any real value,
it is necessary to start with a free particle
on the punctured plane. In this case
the orbital angular momentum operator
is changed for
\begin{equation}
\hat L=-i\frac{\partial}{\partial\varphi}
+\vartheta,\qquad
\vartheta\in \R.
\label{lth}
\end{equation}
Effectively such a change eliminates
the restriction on
the $\alpha$ (for the details, see ref. \cite{PR}).
Note that physically the particle on the punctured plane
with the angular momentum (\ref{lth})
corresponds to the system of a point charged particle
in a field of the singular magnetic flux placed at
$x_i=0$ \cite{Jack,Ler},
and as for the 3D charge-monopole system \cite{GO,mno},
(\ref{lth}) is the total angular momentum
of the particle and electromagnetic field.
Further on we shall suppose that the parameter $\alpha$
can take any real value.

Before passing over to the generalization of the
construction
for the case of nonlinear superconformal symmetry,
we note that the quantum constraint  (\ref{qj1})
can be represented equivalently in the form
\begin{equation}
\hat{\cal J}_1\Psi_{phys}=
\left(\hat{L}+\Pi_+-\alpha -\frac{1}{2}\right)
\Psi_{phys}=0,
\label{pi1}
\end{equation}
where the term $\frac{1}{2}$ is of the quantum origin
(it includes the factor $\hbar=1$),
and $\Pi_+=\frac{1}{2}(\sigma_3+1)$
being a projector,
$\Pi_+^2=\Pi_+$,
is the fermion quantum number
operator,
$\Pi_+=\hat{\psi}_+\hat{\psi}_-$,
$\hat \psi_\pm=\frac{1}{2}(\sigma_1\pm i\sigma_2)$.

\section{Nonlinear superconformal symmetry}
Now we are in position to be able
to generalize the construction for the nonlinear
superconformal symmetry case.
We start, again, from the system of a free
spin-$\frac{1}{2}$ particle
on the (punctured) plane, but instead of (\ref{pi1}),
we postulate the quantum constraint
\begin{equation}
\hat{\cal J}_n\Psi_{phys}=
\left(\hat{L}+n\Pi_+-\alpha -\frac{1}{2}\right)
\Psi_{phys}=0,
\label{pin}
\end{equation}
where $n$ is an arbitrary integer, which for definiteness
is supposed to be positive.
The formal sense of the change of the
quantum condition (\ref{pi1}) for (\ref{pin})
is clear: eq. (\ref{pi1})  singles out the two
eigenstates of $\hat L$ with the eigenvalues
shifted for $1$, while the $\hat L$-eigenvalues
of the upper and lower components of the
spinor satisfying eq. (\ref{pin})  are shifted
relatively in $n$.

Let us demonstrate  that the nonlinear
superconformal symmetry is realized in the
system reduced by the quantum equation
(\ref{pin}). To identify the symmetry generators,
we begin with the
analysis of the corresponding classical system.
The classical analog of the quantum equation
(\ref{pin})
is the constraint
\begin{equation}
{\cal J}_n=L+n\Pi_+ -\alpha\approx 0,
\label{jnclas}
\end{equation}
where $\Pi_+=\xi_+\xi_-$,
$\xi_\pm=\frac{1}{\sqrt{2}}(\xi_1\pm i\xi_2)$,
$\{\xi_+,\xi_-\}=-i$,
and we have omitted the quantum term
$\frac{1}{2}$.
Classically $\Pi_+$ coincides with $\Sigma$,
and so, at $n=1$ the constraint (\ref{jnclas})
takes the form of the constraint (\ref{j1}).
The constraint (\ref{jnclas}) appears
from the Lagrangian
\begin{equation}
{\cal L}_n={\cal L}_0-\frac{1}{2x_i^2}(\epsilon_{jk}x_j
\dot x_k+n\xi_+\xi_--\alpha)^2,
\label{ln}
\end{equation}
with ${\cal L}_0$ given by eq.
(\ref{act}),
as the unique (primary) constraint,
while $H=\frac{1}{2}p_i^2$ is
generated by (\ref{ln})
as the canonical Hamiltonian.

Let us define
the bosonic counterparts
of the mutually conjugate
Grassmann variables $\xi_\pm$,
\begin{equation}
X_\pm=\frac{1}{\sqrt{2}}
(X_1\pm iX_2),\qquad
P_\pm=\frac{1}{\sqrt{2}}(p_1\pm ip_2),
\label{XPdef}
\end{equation}
satisfying the nontrivial Poisson bracket relations
$
\{X_+,P_-\}=\{X_-,P_+\}=1.
$
In terms of these variables, we can identify the
even observables defined as those
having zero Poisson brackets with the constraint
(\ref{jnclas}). These are the same quadratic
quantities (\ref{hkd}), (\ref{lorb}) taking the form
\begin{equation}
H=P_+P_-,\qquad
K=X_+X_-,\qquad
D=\frac{1}{2}(X_+P_-+P_+X_-),
\label{hdk+}
\end{equation}
\begin{equation}
L=i(X_+P_--X_-P_+),\qquad
\Sigma=\xi_+\xi_-.
\label{lxp+}
\end{equation}
To identify the odd observables being integrals
of motion in the sense of eq. (\ref{integr}),
it is sufficient to note that
$
\{\Sigma,\xi_\pm\}=\mp i\xi_\pm,
$
$
\{L,X_\pm\}=\mp iX_\pm,
$
$
\{L,P_\pm\}=\mp iP_\pm.
$
Therefore, the set of odd
independent integrals of motion
commuting with the constraint (\ref{jnclas})
is
\begin{equation}
S^+_{n,l}=2^{n/2}(i)^{n-l}
(P_-)^{n-l}(X_-)^l\xi_+,\qquad
S^-_{n,l}=2^{n/2}(-i)^{n-l}(P_+)^{n-l}
(X_+)^l\xi_-,\qquad l=0,\ldots, n.
\label{sn}
\end{equation}
At n=1 these are the linear combinations of
the odd integrals (\ref{odd1}).

Since on the surface of the constraint
(\ref{jnclas})
the relation
\begin{equation}
{\cal C}\equiv 4(KH-D^2)+2n\Sigma=\alpha^2,
\label{cas}
\end{equation}
is valid, there the quantity ${\cal C}$
commutes
with all the set of the integrals
(\ref{hdk+}), (\ref{lxp+}), (\ref{sn}).
The even integrals (\ref{hdk+})
and $\Sigma$ form, as before, the Lie algebra
$so(1,2)\oplus u(1)$. Then,
treating the constraint (\ref{jnclas})
as that fixing the orbital angular momentum
$L$, and
taking into account the relation (\ref{cas}),
we find that on the surface (\ref{jnclas})
the integrals (\ref{hdk+}),
$\Sigma$ and (\ref{sn})
form the nonlinear superalgebra given
in addition to eq. (\ref{dhk}) by the
following nontrivial Poisson bracket relations:
\begin{equation}
\left\{D,S^\pm_{n,l}\right\}=
\left(\frac{n}{2}-l\right)S^\pm_{n,l},\qquad
\{\Sigma,S^\pm_{n,l}\}=\mp iS^\pm_{n,l},
\label{dsigs}
\end{equation}
\begin{equation}
\left\{H,S^\pm_{n,l}\right\}=
\pm ilS^\pm_{n,l-1},\qquad
\left\{K,S^\pm_{n,l}\right\}=
\pm i(n-l)S^\pm_{n,l+1},
\label{hks}
\end{equation}
\begin{eqnarray}
\left\{ S_{n,m}^{+},S_{n,l}^{-}\right\}=
-i(2H)^{n-m}(2K)^{l}(\alpha -2iD)^{m-l}-i\Sigma
(2H)^{n-m-1}(2K)^{l-1}\times\nonumber\\
\left( \alpha
-2iD\right) ^{m-l}\left(
n\left( m-l\right) \left( \alpha -2iD\right) +4\alpha
l\left( n-m\right)
\right), \qquad m\geq l.
\label{slong}
\end{eqnarray}
The brackets between the odd integrals for the case
$m<l$ can be obtained from (\ref{slong})
by a complex conjugation.
The relations (\ref{dhk}), (\ref{dsigs})-(\ref{slong})
give a nonlinear generalization of the
superconformal algebra $osp(2|2)$
with the Casimir element (\ref{cas}).

In terms of the polar coordinates (\ref{n12}),
(\ref{xpx}),
we have the relations
\begin{equation}
P_\pm=\frac{1}{\sqrt{2}}(p_r\pm ir^{-1}L)e^{\pm i\varphi},
\qquad
X_\pm=\frac{1}{\sqrt{2}}re^{\pm i\varphi} -P_\pm t,
\label{xp+-d}
\end{equation}
while the variables
\begin{equation}
\psi_\pm=\xi_{\pm}e^{\mp in\varphi},\qquad
\{\psi_+,\psi_-\}=-i,
\label{xif}
\end{equation}
are the odd observables commuting with
the constraint (\ref{jnclas}), which
at $n=1$ are transformed to the linear combinations
of the odd variables defined by eq. (\ref{xpx}).
Using the notation $q=r$, $p=p_r$,
and the constraint (\ref{jnclas}),
we obtain the reduced 1D classical Hamiltonian,
\[
H=\frac{1}{2}\left(p^2+\frac{1}{q^2}\alpha(\alpha-2n
\Sigma)\right),
\]
the even,
\[
\Sigma=\psi_+\psi_-,\qquad
D=\frac{1}{2}qp-Ht,\qquad
K=\frac{1}{2}q^2-2Dt-Ht^2,
\]
and the odd,
\[
S^+_{n,l}=2^{n/2}i^{n-l}\left(p-i\alpha q^{-1}\right)^{n-l}
\left(q-\left(p-i\alpha q^{-1}\right)t\right)^l\psi_+,\qquad
S^-_{n,l}=\left(S^+_{n,l}\right)^*,
\]
integrals of motion,
generating the nonlinear generalization
of the superconformal symmetry
$osp(2|2)$. Therefore, the reduction
of the nonrelativistic 2D free
spin-$\frac{1}{2}$ particle system
by the constraint (\ref{jnclas})
produces the 1D classical system of ref. \cite{LP2}
with nontrivially
coupled boson and fermion degrees of freedom,
which possesses the nonlinear superconformal
symmetry.

In ref. \cite{LP2}, it was showed that the quantum
nonlinear
superconformal symmetry is generated by the
set of quantum operators
\begin{equation}
\hat H=\frac{1}{2}\left(
-\frac{\partial ^2}{\partial q^2}+\frac{1}{q^2}\left(
a_n+b_n\sigma_3\right)\right),
\label{hnq}
\end{equation}
with
\begin{equation}
a_n=\alpha_n^2+\frac{1}{4}(n^2-1),\qquad
b_n=- n\alpha_n,\qquad
\alpha_n=\alpha-\frac{1}{2}(n-1),
\label{anq}
\end{equation}
\begin{equation}
\hat \Sigma=\frac{1}{2}\sigma_3,\qquad
\hat D=-\frac{i}{2}\left(q\frac{\partial}{\partial q}
+\frac{1}{2}
\right)-\hat Ht,
\qquad
\hat K=\frac{1}{2 }q^2-2\hat Dt-
\hat Ht^2,
\label{dnq}
\end{equation}
\begin{eqnarray}
&&\hat S{}^+_{n,l}=\left(
q+it{\cal D}_{\alpha-n+1}
\right)
\left(
q+it{\cal D}_{\alpha-n+2}
\right)\ldots
\left(
q+it{\cal D}_{\alpha-n+l}
\right)
{\cal D}_{\alpha -n+l+1}\ldots
{\cal D}_\alpha
\hat\psi_+,\nonumber\\
&&\hat S{}^-_{n,l}=\left(
\hat S{}^+_{n,l}\right)^\dagger,
\label{snq}
\end{eqnarray}
where
\begin{equation}
{\cal D}_\gamma=\frac{\partial}{\partial
q}+\frac{\gamma}{q}.
\label{cdnq}
\end{equation}
The second terms in $a_n$ and $\alpha_n$
in eq. (\ref{anq})
(proportional to $(n^2-1)$ and $(n-1)$)
include the quantum factors $\hbar^2$ and $\hbar$,
respectively,
while the term $\frac{\gamma}{q}$
in (\ref{cdnq}) includes the factor $\hbar(=1)$.
These quantum corrections in the quantum
analogs of the corresponding classical quantities
were found in \cite{LP2} from the requirement of
preservation of the nonlinear superconformal symmetry.
However, their origin remained to be completely unclear.
Now we shall show that the application of the reduction
procedure `first quantize and then reduce' to the system
(\ref{act}), (\ref{jnclas}) produces
exactly the anomaly free
quantum system with the  nonlinear superconformal symmetry
generators given by eqs. (\ref{hnq})--(\ref{cdnq}).

Proceeding from the relations (\ref{n12}),
(\ref{xpx}) and (\ref{XPdef}),
we construct
the quantum analogs of the classical quantities
(\ref{xp+-d}):
\[
\hat P_+=\frac{1}{\sqrt{2}}
\left(
\hat p_r+\frac{i}{r}
\left(
\hat L-{\frac{1}{2}}
\right)\right)e^{i\varphi},\quad
\hat P_-=\left(\hat P_+
\right)^\dagger,\qquad
\hat X_\pm=\frac{1}{\sqrt{2}}r
e^{\pm i\varphi}-\hat P_\pm t.
\]
Putting these relations
into the  quantum analogs of the odd integral
(\ref{sn}),
transporting subsequently the
phase factor $e^{-i\varphi }$
($e^{i\varphi }$ ) in $\hat S{}^+_{n,l}$
($\hat S{}^-_{n,l}$)
to the right (to the left),
defining, in accordance with
relation (\ref{xif}),
the operators
$\hat\psi_+=e^{-in\varphi}\sigma_+$,
$\hat\psi_-=e^{in\varphi}\sigma_-$,
$\sigma_{\pm}=\frac{1}{2}(\sigma_1\pm i
\sigma_2)$,
and passing over to the scalar product
(\ref{newsc})
in the manner described above,
we obtain the fermion operators (\ref{snq}).
The quantum Hamiltonian (\ref{hnq}) is obtained proceeding
from the relation (\ref{hrad})
as before, but now
using the quantum analog (\ref{pin})
of the classical constraint (\ref{jnclas}).
This procedure correctly reproduces
the described quantum corrections
in (\ref{anq}).

In ref. \cite{LP2} the quantum analog of the
nonlinear superconformal algebraic relations
(\ref{slong})
was obtained in a complete form
only for the particular case $n=2$
by a direct calculation of the
anticommutation relations of the
operators (\ref{snq}).
The knowledge of the origin of the
odd operators
(\ref{snq})
allows us to fix the form of
the nonlinear superconformal symmetry
in general case of $n\in\N$.
To this end, we proceed from the
quantum generators
presented in the form
corresponding to the classical expressions
(\ref{hdk+})-(\ref{sn}).
It is obtained via a direct substitution
of the classical quantities
$X_\pm$, $P_\pm$ and $\xi_\pm$
for their quantum analogs satisfying the nontrivial
(anti)commutation relations
\[
[\hat X_+,\hat P_-]=[\hat X_-,\hat P_+]=i,\qquad
[\hat\xi_+,\hat \xi_-]_{{}_+}=1.
\]
Then a simple calculation gives
the following nontrivial commutation relations
\begin{equation}
[\hat H,\hat K]=-2i\hat D,\quad
[\hat D,\hat H]=i\hat H,\quad
[\hat D,\hat K]=-i\hat K,
\label{qso}
\end{equation}
\begin{equation}
[\hat \Sigma,\hat S{}^\pm_{n,l}]=\pm \hat S{}^\pm_{n,l},
\quad
[\hat D,\hat S{}^\pm_{n,l}]=i\left(\frac{n}{2}-l\right)
\hat S{}^\pm_{n,l},
\label{sds}
\end{equation}
\begin{equation}
[\hat H,\hat S{}^\pm_{n,l}]=\mp l\hat S{}^\pm_{n,l-1},\quad
[\hat K,\hat S{}^\pm_{n,l}]=\mp (n-l)
\hat S{}^\pm_{n,l+1},
\label{hks}
\end{equation}
together with the Casimir operator
\begin{equation}
\hat {\cal C}\equiv 2(\hat H\hat K+\hat K\hat H)-
4\hat D{}^2+2n\alpha_n \hat \Sigma  =
\alpha_n^2+\frac{1}{4}n^2-1,
\label{qcas}
\end{equation}
where $\hat \Sigma=\frac{1}{2}\sigma_3$.
To obtain the last relation, we have also used
the quantum constraint (\ref{pin}).
The quantum analog of the classical relation
(\ref{slong})
is given by the anticommutator
\begin{eqnarray}
[\hat S{}^+_{n,m},\hat S{}^-_{n,l}]_{{}_+}=&&
\sum_{s=0}^{min(l,n-m)}2^ss!C^s_{n-m}C^s_{l}
\times (
(2\hat K)^{l-s}(2\hat H)^{n-m-s}
{\cal P}_{m-l+s}(-2i\hat D +c_s)
\Pi_+ +
\nonumber\\&&
(-1)^{m-l}(2\hat H)^{n-m-s}(2\hat K)^{l-s}
{\cal P}_{m-l+s}(
2i\hat D+d_s)
\Pi_-),
\label{ssquant}
\end{eqnarray}
where
$\Pi_\pm=\frac{1}{2}\pm \Sigma$,
$min(a,b)=a$ (or, $b$) when $a\leq b$
(or, $b\leq a$), $C^s_l=\frac{l!}{s!(l-s)!}$,
${\cal P}_k(z)$ is a polynomial of order
$k$,
\[
{\cal P}_0(z)=1,\qquad
{\cal P}_k(z)=z(z+2)\ldots (z+2(k-1)),\quad k>0,
\]
and
\[
c_s=\alpha
+\frac{3}{2}+n-2(m+s),\qquad
d_s=-\alpha +\frac{1}{2}+2(l-s).
\]
In (\ref{ssquant}) we suppose $m\geq l$,
while the case corresponding to $m\leq l$ is obtained from
it
by the Hermitian conjugation.
To calculate the anticommutator (\ref{ssquant}),
we have used the relation
\[
\hat X_+^l\hat P_-^k=\sum_{s=0}^{min(l,k)}i^ss!C^s_lC^s_k
\hat P_-^{k-s}\hat X_+^{l-s},
\]
and the analogous relation with the
$\hat X_+$ and $\hat P_-$
exchanged in their places and with
the $i$ changed for $-i$.
Besides, the product
$\hat X_-\hat P_+=\hat D+\frac{i}{2}(\hat L+1)$
has been presented in the equivalent form
using the relation $\hat L=\alpha +\frac{1}{2}-n\Pi_+$
following from the equation of the quantum constraint
(\ref{pin}).

Eqs. (\ref{qso})--(\ref{hks}),
(\ref{ssquant})
give a general form of the (anti)commutation
relations of the order $n$ nonlinear generalization of the
superconformal algebra $osp(2|2)$, which can be denoted
as $osp(2|2)_n$.
{}From them, in particular, it is
easy to reproduce the simplest nonlinear case
of the $osp(2|2)_2$ superalgebra found in ref.
\cite{LP2}.

\section{Discussion and outlook}
We have showed that the one-dimensional
models
corresponding to the cases of  linear,
$osp(2|2)$,
and nonlinear, $osp(2|2)_n$, superconformal
symmetries can be obtained by the reduction
of the planar free spin-$\frac{1}{2}$ particle system
by the constraint
(\ref{pin}) with $n=1$ or $n>1$.
The reduction produces the nontrivial
coupling of the boson and fermion degrees
of freedom with conservation of the
corresponding (linear or nonlinear)
superconformal symmetry of the initial
system.
This method not only has given a
natural explanation of the origin  of
the quantum corrections
necessarily  to be included in the generators of
the $osp(2|2)_n$ with $n>1$
for preservation of the symmetry at the quantum level,
but also has allowed us to fix the form
of the quantum $osp(2|2)_n$ superalgebra.

The reduction
procedure with the constraint
(\ref{pin}) can
alternatively be treated
as a reduction
of the planar free spin-$\frac{n}{2}$
particle
system. Indeed,
the quantum constraint (\ref{pin})
can be changed for the system of the
quantum equations
\begin{equation}
\left(\hat L +
\hat {\Pi}^{(n)}_+-
\alpha-\frac{1}{2}\right)
\Psi_{phys}=0,
\label{spinn}
\end{equation}
\begin{equation}
\left(\hat\Pi_{1+}-\hat\Pi_{j+}
\right)\Psi_{phys}=0,
\quad
j=2,\ldots, n,
\label{in}
\end{equation}
with
\[
\hat {\Pi}^{(n)}_+=\sum_{k=1}^{n}\hat\Pi_{k+},\qquad
\hat\Pi_{k+}=\frac{1}{2}
\left(\sigma_3^{k}+1\right),
\]
and $\frac{1}{2}\sigma_3^{k}$, $k=1,\ldots,n$, corresponding
to the set of  the $n$ independent
spins of the value $\frac{1}{2}$.
Then eq. (\ref{spinn}) fixes the value of the total angular
momentum,
while the
set of the equations (\ref{in})
prescribes the constituent spins to be polarized in the same
direction.
With taking into account
the quantum constraints (\ref{in}),
the operator in eq. (\ref{spinn})
reduces to the operator
$\hat L+n\hat\Pi_{1+}-\alpha-\frac{1}{2}$,
and after changing the notation $\sigma^1_3\rightarrow
\sigma_3$, eq. (\ref{spinn}) takes a form
of eq. (\ref{pin}).
The corresponding classical Lagrangian for such a
planar system is
\[
{\cal L}^{(n)}=\frac{1}{2}\dot{x}_i^2-\frac{i}{2}\dot{\xi^k}
_i
\xi^k_i
-\frac{1}{2x_i^2}(x_1\dot x_2-x_2\dot x_1+\xi^k_+\xi^k_-
-\alpha)^2+\sum_{j=2}^n v_j(\xi^1_+\xi^1_--\xi^j_+\xi^j_-).
\]
Here $v_j$, $j=2,\ldots n$,
is the set of Lagrange multipliers,
$\xi^k_i$, $k=1,\ldots, n$, is the set
of $n$ planar Grassmann vectors, and
$\xi^k_\pm=\frac{1}{\sqrt{2}}
(\xi^k_1\pm i\xi^k_2)$.

To conclude, we enumerate shortly possible
applications and generalizations of the results.

It would be interesting to generalize  the described method
of introduction of the boson-fermion coupling for other
systems. We hope that this, on the one hand, could
clarify
the nature of the nontrivial quantum corrections appearing
generally under attempt of quantization of the
systems possessing nonlinear supersymmetry
\cite{P1,KP1,KP3};
on the other hand,
the method could be useful
for the analysis
of the quasi exactly solvable systems
\cite{QE1,QE2,QE3},
to which the nonlinear supersymmetry is intimately related
\cite{KP1}--\cite{KP3}.
If the method admits a generalization for higher dimensions,
it could be applied  to investigation of
the supersymmetric many-particle integrable systems.

\vskip 0.3cm
{\bf Acknowledgements}
\vskip 3mm
The work has been supported in part by
the FONDECYT-Chile (grant 1010073).

\end{document}